\documentclass[aps,prl,superscriptaddress, twocolumn,%showpacs,
preprintnumbers,amsmath,amssymb]{revtex4}
\usepackage{color}
\usepackage{array}
\usepackage{amsmath}
\usepackage{amssymb}
\usepackage{epsfig}
\usepackage{graphicx}
\usepackage{wasysym}
\usepackage{bbm}

\begin{document}
\title{Composite Fermions and Quantum Hall Stripes on the  Topological Insulator Surface}
\author{Ying Ran}
\affiliation{Department of Physics, Boston College, Chestnut Hill, MA 02467}
\author{Hong Yao}
\author{Ashvin Vishwanath}
\affiliation{Department of Physics,
University of California at Berkeley, Berkeley, CA 94720, USA}
\affiliation{Materials Sciences Division,
Lawrence Berkeley National Laboratory, Berkeley, CA 94720, USA}

\date{\today}

\begin{abstract}
We study the problem of a single Dirac fermion in a quantizing orbital
magnetic field, when the chemical potential is at the Dirac point.
This can be realized on the surface of a topological insulator, such
as Bi$_2$Se$_3$, tuned to neutrality. We study the effect of both
long range Coulomb interactions (strength
$\alpha=\frac{e^2}{\epsilon\hbar v_F}$) and local repulsion $U$
which capture the effect of electron correlations. Interactions resolve the degeneracy of free fermions in the zeroth Landau level at half filling, but in a manner different from that in graphene. For weak interactions, $U=0$ and $\alpha \ll 1$, a composite Fermi liquid is expected. However, in the limit of strong local
correlations (large $U$ but $\alpha \ll 1$), a charge density wave phase is predicted, which we term
``axion stripe''. While reminiscent of quantum Hall stripe phases, its
wavelength is 
parametrically larger than the magnetic length, and
the induced fermion mass term (axion) also oscillates with the charge density. This phase is destroyed by sufficiently strong Zeeman
coupling. A phase diagram is constructed and consequences for
experiments are discussed.
\end{abstract}

\maketitle

There has been much recent interest in topological insulators (for reviews see \cite{hasan_kane_review,qi:33}) with
strong spin-orbit interactions, which feature protected surface
states in the presence of time-reversal invariance (TRI). A minimal
model of the surface states of a three dimensional strong
topological insulator(TI) is a single Dirac node, which is impossible to
obtain in a 2D band structure with TRI. Photoemission experiments on
e.g. Bi$_2$Se$_3$ reveal precisely such a surface band. These surface
states may be considered as $1/4$th of graphene, the other well
known Dirac metal, which has additionally two fold valley and spin 
degeneracies.

A key feature of Dirac metals is their anomalous response to a
magnetic field, which includes large Landau level (LL)
spacing and an
unusual offset in quantum Hall states. For a single Dirac node at
neutrality (chemical potential at the Dirac point), the zeroth
LL is half filled, implying an enormous degeneracy in the
clean limit. Interactions are critical to lifting this degeneracy.
The precise nature of the resulting state is the subject of this
paper. When graphene is tuned to neutrality in a magnetic field,
this Landau degeneracy is resolved by spontaneous symmetry breaking
- two of the four LLs are chosen to be completely filled
and a translationally invariant 
state is obtained. In the case here
with a single Dirac node, this is not an option - the only ways
for the system to lift the degeneracy is by forming a homogeneous fractional quantum hall (FQH) phase, or breaking the translational symmetry
to form spatial patterns as in the stripe phase. Recently, Dirac
like LLs on the surface of Bi$_2$Se$_3$ were
observed \cite{cheng-2010,Hanaguri-2010}. Although the chemical potential there was not
tuned to neutrality, and the LL widths indicate the
importance of disorder, this is a hopeful sign for future
experimental work along the direction of this paper.

The short range interactions of the TI surface modes will be
parameterized by $U$, and the long range Coulomb interactions by the
dimensionless coupling $\alpha\equiv e^2/(\epsilon \hbar v_F)$. 
Here $v_F$ is the 
fermi velocity and
$\epsilon=\frac{1+\epsilon_{bulk}}{2}$ the effective surface dielectric
constant. We will assume $\alpha\ll1$ throughout, which allows us to
treat the Coulomb interactions last. This is %a 
reasonable in currently known TIs
since $\epsilon \gg 1$ (e.g.,
$\epsilon_{bulk}=290$ \cite{BiSe_dielectric} in Bi$_2$Te$_3$) while $v_F\sim 3\times 10^5
{\rm m/sec}$ in Bi$_2$Te$_3$ \cite{Chen07102009, Hsieh_Tunable}, thus $\alpha$ $\sim$
$10^{-1}$ to $10^{-2}$. Estimating $U$ is hard, but it is
expected to be large in more correlated systems like the Iridates \cite{PhysRevLett.102.256403, pesin-2009}. 

\begin{figure}
\includegraphics[width=0.22\textwidth] {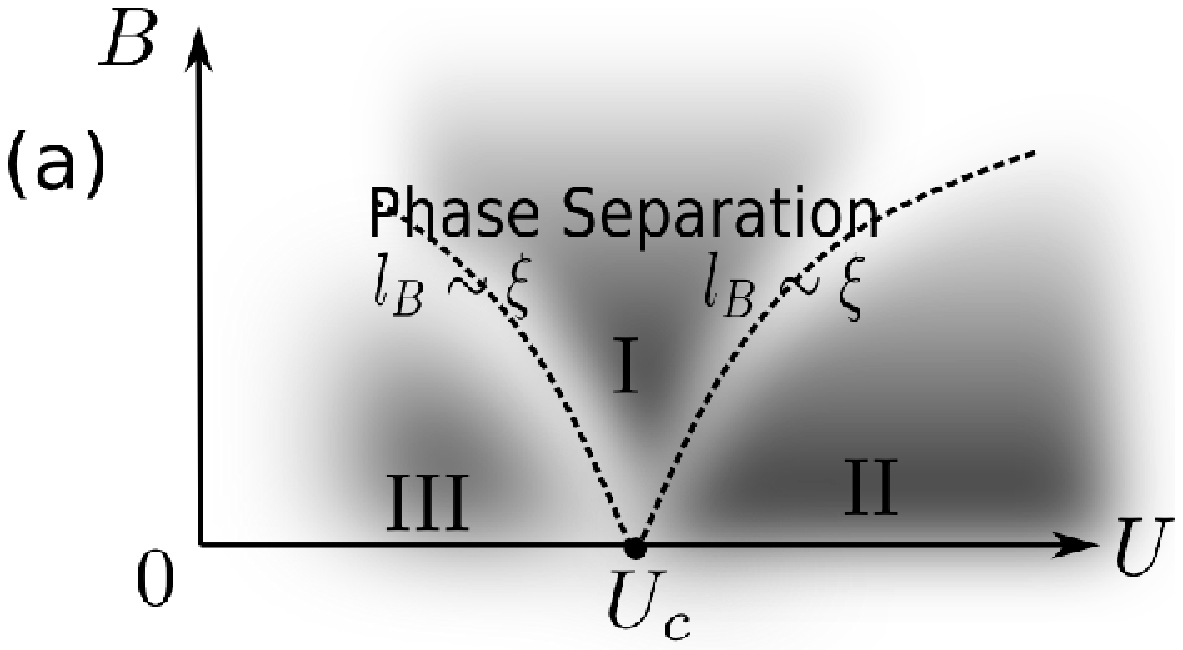}
\includegraphics[width=0.22\textwidth] {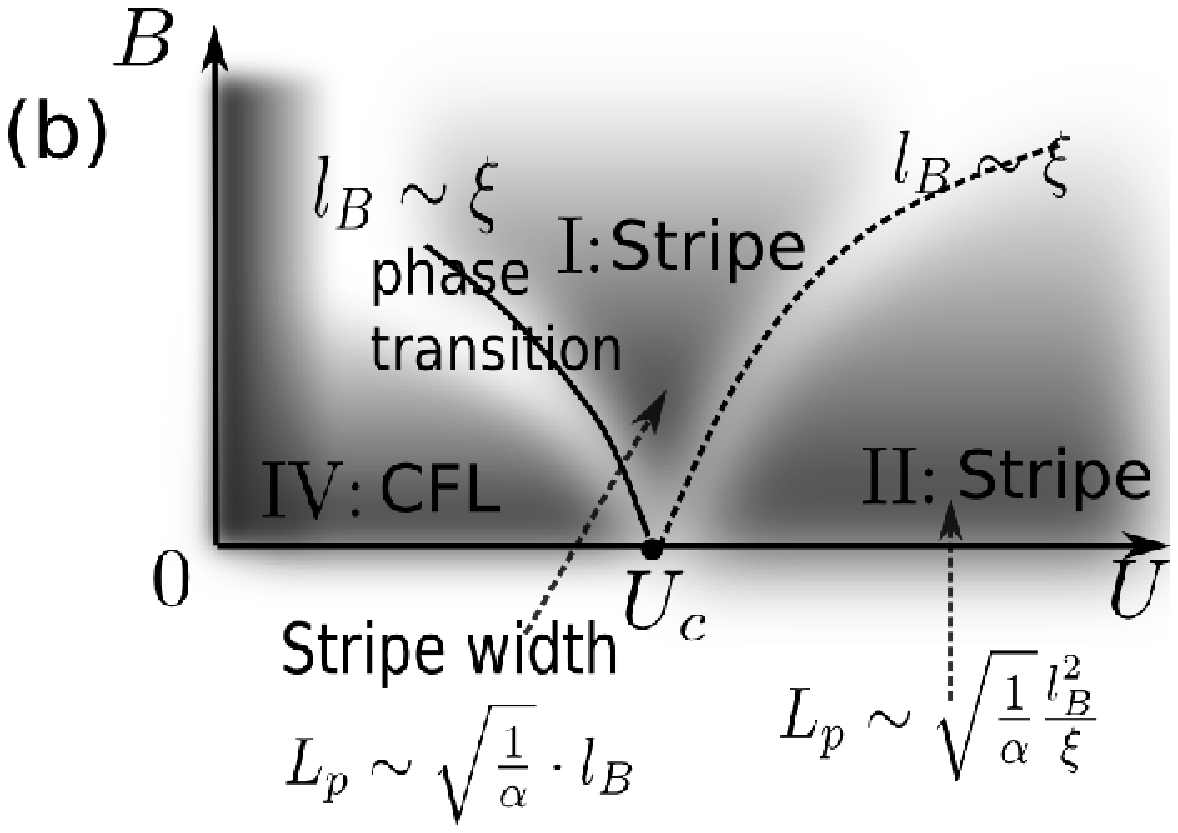}\\
\includegraphics[width=0.35\textwidth] {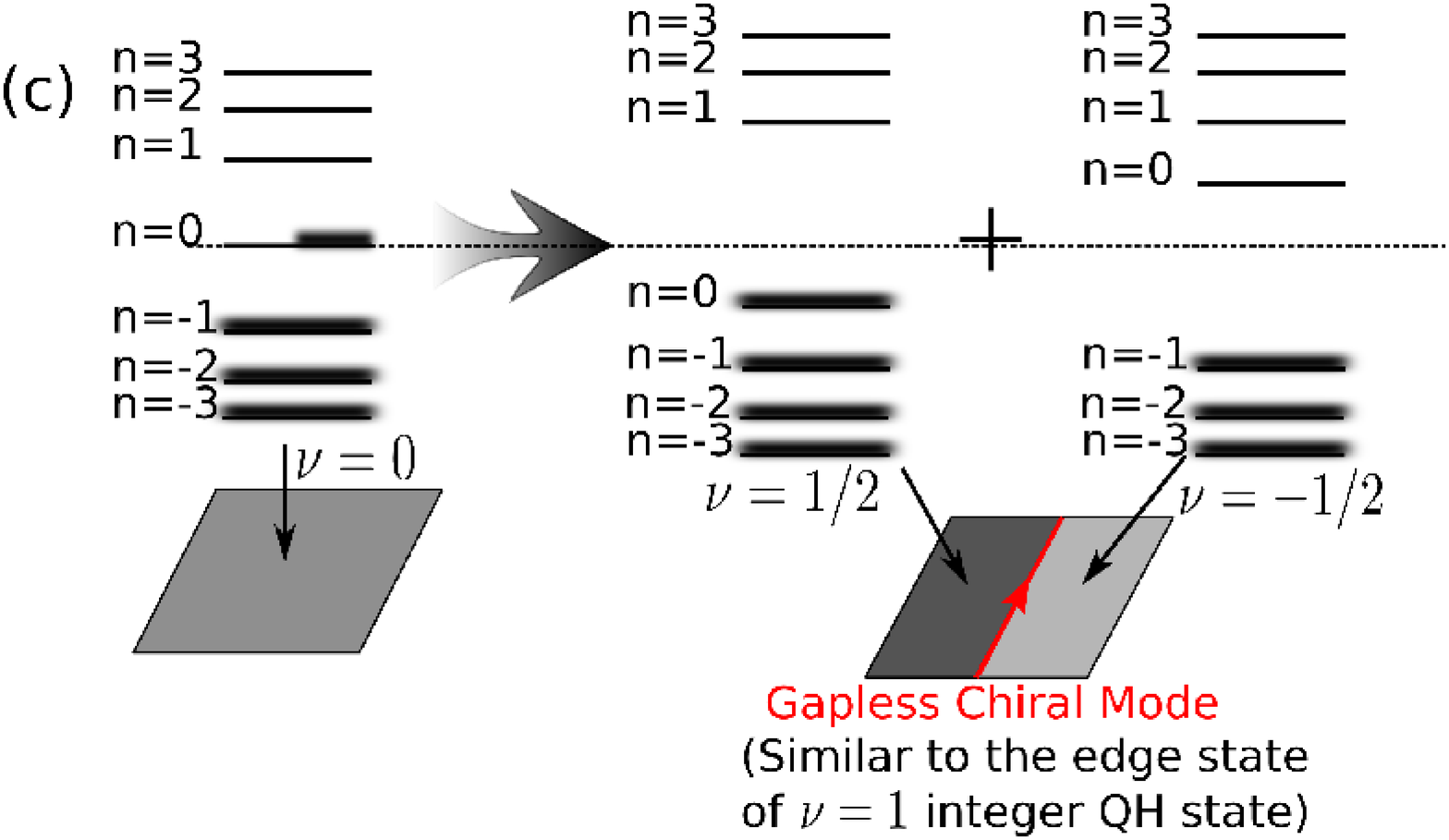}
\caption{ Zero-temperature phase diagram of a single Dirac node, in
an external magnetic field, with a tunable short-range repulsion $U$
and (a) no long-range Coulomb interaction, or (b) with a fixed weak
long-range Coulomb interaction $\alpha\ll 1$. In both (a) and (b),
we assume the chemical potential is at the Dirac node. Region I:
quantum critical regime with $l_B\ll \xi$. Region II: weak field
limit in the magnetic ordered phase, namely magnetic length $l_B$ is
much greater than any intrinsic length scale.
Region III: weak field limit when
$l_B\gg \xi$ and $U \lesssim U_c$. Region IV: Weak field limit
when $l_B\gg \xi$ and $U \lesssim U_c$, and also extended over the
whole weak $U$ regime. Phases predicted in regime I and III are valid in
large-$N$ limit. Phases predicted in regime II and IV are valid
for arbitrary $N$, including $N=1$ (allowing only for flavor symmetric
states). (c) Schematic picture of
phase separation realized in region (a).}
\label{fig:phase_diagram}
\end{figure}

In zero magnetic field, when the surface is at neutrality
(recent experiments reported control over the surface chemical potential \cite{PhysRevLett.103.246601}), a metal to
insulator transition is expected when $U$ is varied across a critical value $U_c$, 
due to the spontaneous generation of a magnetic mass term $m$. In an orbital magnetic field, our
results are as follows. When $\alpha=0$, we find within
a Hartree-Fock (large-$N$) approximation that phase separation occurs for the entire
range of $U$, into a pair of domains with completely filled and
empty zeroth Landau level. On introducing weak Coulomb interactions
$\alpha \ll 1$, macroscopic phase separation is forbidden \cite{Emery1993597}. In the following two regimes the
resulting ground state may be predicted reliably. 
For $U\ll U_c$, a homogenous phase, the Halperin-Lee-Read composite Fermi liquid(CFL) is expected. In the opposite limit, for $U>U_c$, a charge density wave
(stripe) is obtained where the mass term ``$m$'' is also modulated. We term it as ``axionic stripe'' phase because in
contrast to usual quantum Hall stripes\cite{PhysRevLett.76.499,PhysRevB.54.5006,PhysRevLett.84.1982}, the wavelength $\lambda$ here is
parametrically different from $l_B$, the magnetic length.
Indeed, 
$\lambda \sim l_B^2{(\sqrt{\alpha}\xi)^{-1}}$, where $\xi$ is the
correlation length in the insulating phase. 
Other possible intermediate phases are also discussed. The results are summarized
in Figure \ref{fig:phase_diagram}. The Zeeman coupling, when small,
does not modify the phase diagram. However, if it is sufficiently
large, it can destroy the "axion stripe" phase. The critical Zeeman
coupling (or $g$ factor: $g=g_c$) depends on microscopic details
such as the ultraviolet cutoff, but typically $g_c \sim {\rm O(1)}$.
Finally, we point out that coating the available TI surfaces with a
magnetic layer could also lead to a stripe state.

{\em Model with short-range repulsion}: The effective theory of 
the surface states in an external magnetic field is:
\begin{eqnarray}
 {\cal H}=
\psi^{\dagger}_\alpha\big[\vec \Pi\cdot\vec \sigma\big]_{\alpha\beta} \psi_\beta+U(\psi^{\dagger}_\alpha\psi_\alpha)^2,
\label{eq:fullH}
\end{eqnarray}
where ${\cal H}$ is hamiltonian density, $\vec \Pi=\vec
p+\vec A$, $U=Ua^2$ is the short-range repulsion when $a$ lattice
constant set to one for simplicity (this is the only possible four-fermion term without derivative for a single Dirac fermion), and $\vec \sigma=(\sigma_x,\sigma_y)$ spin Pauli matrices.
In this paper, we set $v_F$, $\hbar$, $e$, and $c$ to  one unless stated otherwise. Note that Zeeman coupling is ignored above but will be
included later. For $B=0$ with weak interactions a
Dirac semi-metal is obtained, since $U$ is irrelevant by power
counting. A
metal-insulator transition is expected when $U>U_c$. A unique
feature of the current system is that the insulating phase must
break time-reversal symmetry, namely a magnetic mass term $m\psi^{\dagger}\sigma_z\psi$ is
generated. Such a mass generation mechanism, if continuous, is
described by the Gross-Neveu universality \cite{PhysRevD.10.3235}. We first study this
phase transition and the magnetic field effect at the Hatree-Fock (or mean-field)
level, which is essentially equivalent to a large-$N$ approximation(see suppl. mat.) By the mean-field (MF) approximation,
\begin{eqnarray}
H_{MF}=\psi^\dag[\vec \Pi\cdot\vec \sigma+ m\sigma_z]\psi +\frac{m^2}{2U}+U\frac{\rho^2}{2}
\end{eqnarray}
where $m$ and $\rho$ are the variational parameters;  $\rho=\langle\psi^{\dagger}\psi-1\rangle$ 
and $m=-U\langle\psi^{\dagger}\sigma_z\psi\rangle$ are the self-consistent conditions to minimize energy. Note that $\rho$ is the electron density with respect to half-filling. 

For a homogeneous $m$, the energies
of LLs are:
\begin{align}
\epsilon_n=\begin{cases}\mbox{sgn}(n)\sqrt{2|n|B+m^2},&\mbox{ if }n\neq0\\
    -m &\mbox{ if }n=0,
\end{cases}
\end{align}
where $B>0$ (along positive $z$ direction) is assumed. LL degeneracy is $\frac{B}{2\pi}$. Close to charge neutrality all the negative LLs are fully filled and the zeroth LL is partially filled. $\rho=\frac{B}{4\pi}$ ($-\frac{B}{4\pi}$) for the fully filled (empty) zeroth LL. % respectively. 
The MF ground state energy is: 
\begin{align}
&E(\rho,m;B)=-m\rho+\frac{B|m|}{2\pi} -\frac{(2B)^{3/2}}{4\pi}\zeta(-\frac{1}{2},\frac{m^2}{2B})\notag\\
& \qquad\qquad\qquad -\frac{(m^2+\Lambda^2)^{3/2}}{6\pi}
 +\frac{1}{2U}m^2+\frac{U\rho^2}{2},
\label{eq:vari_mf}
\end{align}
where $\zeta$ is the generalized Riemann-Zeta function.   
The derivation is in the supplemental material. Similar results have been obtained in Ref. \cite{Zhukovsky:February2003:0040-5779:254}. We point out that the coefficient of the second term in our result is twice of the previous incorrect result, which is essential to obtain the right phase diagram. Here we explain the physical consequences. For
$B=0$ and $\rho=0$, the quadratic term is $\delta m^2$ with
$\delta=1/(2U)-\Lambda/(4\pi)$, from which we obtain the critical
value $U_c=2\pi/\Lambda$ [corresponding on-site repulsion on a lattice is $2\pi/(\Lambda a^2)$], at zero
field. The mass correlation length $\xi$ is $\xi \sim 1/\delta$. For
finite $B$, the key feature of Eq. \ref{eq:vari_mf} is the presence
of the linear term $-m\rho$, whose form is dictated by the 
symmetry argument below.

Consider the time-reversal (TR) symmetry and the (unitary)
particle-hole (PH) symmetry defined by $\psi_\alpha
\stackrel{\mathrm{PH}}{\longrightarrow}
[\sigma_x]_{\alpha\beta}\psi^\dagger_\beta$. The massless Dirac
theory as well as the interactions respects both of them. Even though
external magnetic field $B\rightarrow -B$ under each of these, it
respects the combination $\mathrm{TR}\circ \mathrm{PH}$, namely $B\stackrel{\mathrm{TR}\circ\mathrm{PH}
}{\longrightarrow}B$. 
Since $m\stackrel{\mathrm{TR}\circ\mathrm{PH}
}{\longrightarrow}-m$ and $\rho \stackrel{\mathrm{TR}\circ\mathrm{PH}
}{\longrightarrow}-\rho$, we conclude that $E(\rho, m; B)=E(-\rho,
-m; B)$, 
which dictates that the only
possible linear term is $- m\rho$. 

The linear term $-m\rho$ has important consequences. At
half-filling, $\rho=0$ everywhere if the system is homogeneous; 
consequently the system cannot take advantage of the linear term to gain
energy. To have a finite $\rho$ to gain energy from the linear term,
the system can break translational symmetry either microscopically by
developing a density wave or macroscopically by phase separation.
The latter is realized with short-range interactions, but we will
see below that on including the long range Coulomb repulsion, a
density wave can result.
Note, since $\mathrm{TR}\circ\mathrm{PH}$ is a symmetry of the Hamiltonian [Eq.
(\ref{eq:fullH})] at half-filling, phase separation or a charge density wave(CDW) which involves
generating a finite $m$ or $\rho$ spontaneously breaks the TR$\circ$PH symmetry.

We denote the optimal mass as $m^\ast$ if the system is homogeneous {\it i.e.} $\rho=0$ everywhere. From Eq. \ref{eq:vari_mf}, $m^\ast=0$ for $U<U_c(B)=[U^{-1}_c-| \zeta(\frac12)|(2B)^{1/2}/(4\pi)]^{-1}$. To see whether the system tends to  phase separate into two regions with $\pm ( m,\rho)$, we examine the variational energy Eq. (\ref{eq:vari_mf}) around $m^\ast=0$ by setting $\rho= m/U\ll 1$: 
\begin{eqnarray}
E(\rho, m;B)-E(0,0;B)=-\left[ \frac{(2B)^{1/2}}{8\pi}\zeta(\frac{1}{2}) +\frac{\Lambda}{4\pi}\right] m^2, 
\end{eqnarray}
where higher order terms in $m$ has been neglected. 
We find that phase separation occurs when $a\lesssim l_B$, which is always satisfied for currently
available magnetic fields. For $U>U_c(B)$, $m^\ast\neq 0$. In this case, the system always has phase separation tendency because the energy gain by forming $\pm (m^\ast,\rho)$ is $-|m^\ast \rho|$ which dominates energy cost $U\rho^2/2$ for finite but small enough $\rho$. What is the optimal value of $\rho$ to minimize the energy? It
is tedious but straightforward to show that the optimal $\rho=\pm \frac{B}{4\pi}$, namely the
zeroth LL is either fully filled or empty. Thus the two
domains are quantum Hall states with $\nu=\pm\frac12$, with a chiral
edge mode between them.

Below we demonstrate that in some regimes phase separation is reliable even beyond the MF level.
First, in region II of Fig. \ref{fig:phase_diagram}(a), we are in massive phase with a weak magnetic field. Let
us denote the optimal mass as $m_{*}=\pm |m_*|$. Due to the linear term
$-\rho m_*$, the phase separation energy gain is $-|m_*|
\frac{B}{4\pi}$, which dominates any
other regular $B^2$ energy cost for small enough $B$. 
Second, in regime III, $U\lesssim U_c$ and magnetic field is weak. The Renormalization Group (RG) flows to the free Dirac fermion fixed point (FP) where perturbative treatment is reliable. This dictates that the energy gain of phase separation in region III scales as $\sim -\xi B^2$,  where $\xi$ is the correlation length at zero magnetic field that diverges at criticality. This again dominates any regular $B^2$ cost when $U$ close to $U_c$.  Finally in the quantum critical regime (I), 
$l_B\ll \xi$, $m$ 
$\sim B^{1/2}$ and the energy gain 
$\sim -B^{3/2}$ which also dominates sub-leading regular $B^{2}$ term. The arguments of phase separation for regime II only rely the gaussian massive FP, which is valid for arbitrary $N$ including $N=1$. The sign of the phase separation energy gain in regimes I,III, however, is only reliable in the large-$N$ limit. (see suppl. mat.)

{\em Coulomb Interactions:} What happens if a weak long-range Coulomb interaction is included? First, macroscopic phase separation
is no longer possible \cite{Emery1993597} because of the infinite charging energy cost.
Thus, either a homogeneous system or alternating spatial patterns of
$\nu=\pm1/2$ regions with characteristic pattern size
$l_p$ will result. If eventually $l_p\gg l_B,\xi$, our previous
calculations can still be trusted within each region, which allows
us to treat Coulomb interaction as a perturbation classically:
$H_{\mathrm{Coulomb}}=\frac{e^2}{2\epsilon}\int
dr^2dr'^2\frac{\langle\rho(r)\rangle\langle\rho(r')\rangle}{|r-r'|}$.
We show that this is indeed the case in regime I and II.

By dimensional analysis the Coulomb energy density cost is
$\sim \frac{l_p}{\epsilon l^4_B}$. The energy density cost of the domain
wall separating different regions is $\gamma/l_p$
where $\gamma$ is the domain wall tension. By minimizing the total
energy cost, the optimal $l_p$ is found to be $l_p\sim
\sqrt{\epsilon\gamma}l^2_B$ with the optimal energy density cost %per unit area 
$\sim \sqrt{\frac{\gamma}{\epsilon}}/l_B^2$. Note that this
dimension analysis is independent of the exact shape of patterns, which
will be considered shortly. 

In quantum critical regime (I),
$\gamma\sim 1/l_B^2$ by hyperscaling hypothesis. Thus, $l_p\sim
\sqrt{\epsilon}l_B\gg l_B$ when $\epsilon=1/\alpha\gg 1$. The total
energy density cost is $\sim \sqrt{\alpha}/l_B^3$, much smaller than phase separation gain $\sim 1/l_B^3$. In regime II, $l_B\gg \xi$. When $U$ is close to $U_c$, $\gamma\sim 1/\xi^2$ by hyperscaling hypothesis and the resulting optimal pattern size $l_p\sim \sqrt{\frac{1}{\alpha}}\frac{l_B^2}{\xi}\gg l_B$.  The optimal energy density cost $\sim \sqrt{\alpha}/(l_B^2 \xi)$ is again much less than the phase separation gain $\sim 1/(l_B^2 \xi)$. Deep in the ordered phase $U\gg U_c$, $\sqrt{\gamma}$ and $m$ are high energy scales where short-range physics is important. Their ratio is a non-universal number which generically is of order one. Comparing energy cost $\sim \sqrt{\alpha\gamma}/l_B^2 $ and gain $\sim m/l_B^2$ it is clear that the ground state is also in stripe phase when $\alpha \ll 1$.

\begin{figure}
 \includegraphics[width=0.25\textwidth]{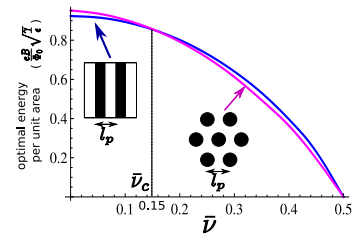}
\caption{(color online) Variational results of the most likely
patterns: Stripe (blue) and triagular lattice CDW (magenta). In the triangular lattice CDW phase we use the simple
ansatz where the negative mass domains (black regions) form circles.
Close to half-filling the optimal pattern is stripe.}
\label{fig:pattern}
\end{figure}

What is the spatial pattern? The most natural options are stripes or
a triangular lattice similar to a Wigner crystal. 
As this is
completely  
determined by the classical Coulomb interaction, we
perform a variational calculation with $H_\mathrm{Coulomb}$ and domain wall
tension $\gamma$. We work in thin domain wall limit (domain wall
width $\ll l_p$) and carry out numerical Ewald-Summation to
determine the precise coefficient in the Coulomb energy density $\sim \frac{l_p}{\epsilon l_B^4}$ for each trial ansatz. The coefficient in the domain wall energy density $\sim \frac{\gamma}{l_p}$ is determined by geometry straightforwardly. The optimal $l_p$ and energy cost is found by minimizing the sum of the two. As a function of average 
$\bar\nu$, we find the phase diagram in
Fig.\ref{fig:pattern}. When $\bar \nu \approx 0$ the optimal pattern
is stripe; when the charge unbalance (the area unbalance between
$\nu=1/2$ and $\nu=-1/2$ regions) is increased, a first order
transition is found at $\bar\nu\approx\pm 0.15$ into the triangular
lattice charge density wave (CDW). (A narrow region of intermediate phases is expected around this putative first order transition\cite{Spivak20062071}.) This filling dependence of the CDW 
pattern
is similar to the one in the high LLs of a parabolic band \cite{PhysRevLett.76.499,PhysRevB.54.5006,PhysRevLett.84.1982}.

Finally we discuss regime IV in Fig.\ref{fig:phase_diagram}(b). In this weak field limit the short range $U$ is irrelevant but Coulomb interaction is marginal. Thus the Coulomb interaction always dominates and we can ignore $U$.
If LL mixing is ignored when $\alpha\ll 1$, because the wave functions in the zeroth LL of a Dirac metal are the same as the lowest LL of a quadratic band (see suppl. mat.), the same set of Haldane pseudo-potentials is obtained for both cases, indicating the ground state in a CFL phase as confirmed by exact diagonalization numerics \cite{PhysRevLett.84.4685}.

What happens if LL mixing is included? This is a rather neat problem because LL mixing is independent of magnetic field and only controlled by dimensionless constant $\alpha$. 
We perform a standard second-order perturbation study by integrating out all the nonzero LLs in a single step, which gives corrections to the interaction in the zeroth LL. (See suppl. mat. Similar study has been performed on the LL mixing of quadratic band \cite{PhysRevB.80.121302}.)
We find the correction to pseudo-potential in the 0th LL is:
\begin{align}
V_1&=0.443-0.0068\alpha, &V_3&=0.277+0.0028\alpha.
\end{align}
This is a very small effect. It is worth noting that the direction of the first order correction is towards a Pfaffian phase and a stripe phase eventually \cite{PhysRevLett.84.4685}.

{\em Zeeman Splitting:} We comment on the effect of Zeeman coupling which we have ignored so far. For the 0th LL, its electron spin is fully polarized along $-z$ direction such that its Zeeman energy is a constant no matter how its electrons are distributed spatially. The Zeeman coupling is only through higher LLs. As shown in the suppl. mat., with only short-range $U$, phase separation is still favored in the MF approximation as long as $U$ is not too strong compared with $U_c$. On introducing the long range Coulomb
interaction however, the ``axionic stripe'' phase in regime II may
be destroyed by the Zeeman coupling. A simple estimate of Zeeman
energy density is $g\mu_B B\langle
\psi^{\dagger}\sigma_z\psi\rangle=g\mu_B B\frac{m}{U}$. Comparing
with phase separation energy gain $\frac{mB}{4\pi}$ one finds that
the condition for the ``axionic stripe'' phase to survive is
$\frac{4\pi\hbar g\mu_B}{e U}< 1$ (we put back units. $U>U_c$ and
$U_c\sim \hbar v_F /\Lambda$). This ratio is material dependent and
can be of order unity (see suppl. mat.).  Thus, we conclude that for
materials with small $g$ and/or large $v_F,a$, Zeeman energy is only
a small and negligible perturbation in the full phase diagrams of
Fig. \ref{fig:phase_diagram}. On the other hand if $\frac{4\pi\hbar
g\mu_B}{e U_c}> 1$, the ``axionic stripe'' phase in regime II of
Fig.\ref{fig:phase_diagram}(b) will be destroyed but other regimes
are still valid. Which phases would then take its place is presently
unclear, and left to future work.

The short range $U$ is likely to be smaller than $U_c$ in currently known TIs, which are then likely in a CFL phase when the fermi level is close to the Dirac node and external  magnetic field is applied. How to realize the predicted stripe phase? One way is to search for TIs with relatively strong $U$ such that $U\gtrsim U_c$; another way is to coat the surface with ferromagnetic (FM) materials such as Mn or Fe while keeping the surface states close to half-filling. For the latter, a magnetic mass is induced on the surface states of TI via exchange. Then the system can be effectively treated as in regime II and the stripe phase is preferred as a ground state when Zeeman coupling is small. The stripes and associated chiral edge modes along domain walls can be directly detected by microscopic probes such as STM.

At what temperature can those predicted phases in Fig.\ref{fig:phase_diagram} be realized? For the CFL phase in regime IV, its signature of  can be detected only if the temperature $k_B T\ll \frac{e^2}{\epsilon l_B}$, which is quite low given the large $\epsilon$ of the current systems. For macroscopic phase separation, they have a lower entropy density compared with the half-filled LL and the difference can be estimated to be $k_B \frac{B}{2\pi} \ln 2\sim \frac{k_B}{l_B^2}$ by counting the number of choices to half fill the zero LL. 
Thus the phase separation critical temperature $T_c\sim \frac{E_\mathrm{gain}l_B^2}{k_B}$ where $E_\mathrm{gain}$ is the phase separation energy density gain. In the quantum critical regime I: $T_c\sim \frac{\hbar v_F}{k_B l_B}$ which is of same order of the LL spacing. In regime II: $T_c\sim \frac{\hbar v_F}{k_B \xi}$ when $U \gtrsim U_c$ and $T_c$ becomes a high-energy scale $\sim \Lambda$ in the deeply ordered phase. We conclude that phase separations in regimes I and II are very stable to finite temperature. Finally for the weak field phase separation regime III: $T_c\sim \frac{\hbar v_F\xi}{k_B l_B^2}$. With Coulomb interactions, if underlying lattice is ignored, the stripe phases breaks
 continuous translational symmetry which would be 
restored in 2D at any finite temperature. However, 
because of the lattice as well as disorder, we expect 
that stripes is pinned and become static which is 
amenable to STM investigations\cite{Spivak20062071}.

{\em Conclusion:} We have argued for the existence of a composite
fermi liquid at small $U$ and a stripe phase for large
$U$, in the presence of Coulomb interactions and a
quantizing orbital magnetic field. The phase structure for intermediate
$U$ is difficult to predict reliably, although large-$N$ calculations predict a stripe phase there as well. This regime deserves
further study, since it could  harbor interesting homogeneous phases, like
the Read-Moore state which is natural at this filling, with large
characteristic energy scales. We hope these results will
encourage experiments and the search for topological
insulators with strong correlations.

We thank Dung-Hai Lee, Steve Kivelson, and Jason Alicea for helpful discussions. YR is supported by the start-up fund at Boston College. HY and AV are supported by
DOE grant DE-AC02-05CH11231.

\newpage 
\begin{center}
 {\bf Supplemental Material}
\end{center}

\appendix
\section{Charge neutral point of an inversion symmetric perfect topological insulator}
In the majority of our study we assume the chemical potential is at or close to the Dirac node of the surface state. This is certainly possible if the sample is gated or doped so that the chemical potential is tunable. Here we try to further point out that for a perfect sample of a topological insulator \emph{with inversion symmetry}, which is actually respected by all the current available experimental systems of topological insulators such as Bi$_{1-x}$Sb$_{x}$, Bi$_2$Se$_3$ and Bi$_2$Te$_3$ have inversion symmetry, our assumptions is automatically realized.

Consider a slab of topological insulator with open boundary condition along the $z$-direction. One can view a whole chain of unit cells along the $z$-direction as a single enlarged unit cell. In terms of these enlarged unit cells the system becomes two dimensional whose fermi surface states are those surface states. By inversion symmetry the top surface and bottom surface fermi surface volumes must be identical. And by Luttinger's theorem band insulators should have total fermi surface volume zero. This dictates only two possibilities: the surface fermi volume is zero or one-half of the Brillouin Zone. The second case is more subtle and related to an additional topological index of the polyacetylene-type: The edge states of polyacetylene chains along $z$-direction has electron charge $\pm 1/2$. Therefore the additional band forms a fermi surface volume of half the surface Brillouin Zone at half-filling. In the present study we assume the TI has trivial polyacetylene-type index and chemical potential is at Dirac node.

\section{About Eq. (\ref{eq:vari_mf})}
Eq.(\ref{eq:vari_mf}) plays a key role in our study and we will 
prove its validity in this section. First, as far as we are aware of, the correct form of the energy function is written down as in Eq. (\ref{eq:vari_mf}) for the first time. 
The major difference between the current result and the previous results is that, when $\rho=0$, the energy has no linear dependence of $m$. The previous results \cite{Zhukovsky:February2003:0040-5779:254} are incorrect because the regularization scheme is wrong. 

Now we derive Eq. (\ref{eq:vari_mf}). In an external magnetic field, we consider the orbital effect and ignore the Zeeman coupling for the moment since adding its contribution to energy function form is straightforward. The Hamiltonian in an external magnetic field is given by 
\begin{align}
 H&=v_F \psi^{\dagger}\begin{pmatrix}
        0&\Pi_x-i\Pi_y\\
\Pi_x+i\Pi_y&0
       \end{pmatrix}\psi+U(\psi^{\dagger}\psi)^2\notag\\
&=H_0+H_{int}.\;\;\;\;\;\;\;\;\;\;\;\;\;\;\;\;\;\;\;\;\;(\vec\Pi\equiv\vec p+\frac{e}{c}\vec A)
\end{align}
We abbreviated spin indices. Without loss of generality we always assumes $B>0$ along $z$-direction. The interaction can be decoupled into two ways by Wick's theorem:
\begin{align}
\langle H_{int}\rangle_{MF}=U\big(\langle\psi^{\dagger}\psi\rangle+\frac{1}{2}\langle\psi^{\dagger}\psi\rangle^2-\frac{1}{2}\langle\psi^{\dagger}\sigma_z\psi\rangle^2\big),\label{eq:decoupling}
\end{align}
We perform the variational mean-field study and use \emph{homogeneous} trial wave function which is the ground state of:
\begin{align}
 H_{MF}=H_0+m\psi^{\dagger}\sigma_z\psi+\frac{m^2}{2U}+U\big(\rho+\frac{\rho^2}{2}\big),
\end{align}
where $m$ and $\rho$ are the variational parameters;  $\rho=\langle\psi^{\dagger}\psi-1\rangle$ 
and $m=-U\langle\psi^{\dagger}\sigma_z\psi\rangle$ are the self-consistent conditions to minimize energy. Note that $\rho$ is the electron density with respect to half-filling. 

$H_0+m\psi^{\dagger}\sigma_z\psi$ gives the well-known Landau levels (LL) of a massive Dirac cone. The energies of LLs are:
\begin{align}
 E_n=\begin{cases}\mbox{sgn}(n)\sqrt{2|n|\frac{\hbar^2 v_F^2}{l_B^2}+m^2},&\mbox{ if }n\neq0\\
    -m &\mbox{ if }n=0.
\end{cases}\label{eq:ll_energy}
\end{align}
Define $\langle H_0+m\psi^{\dagger}\sigma_z\psi\rangle|_{B=0}=\int_0^{\Lambda} -\frac{\sqrt{p^2+m^2}dp}{2\pi}=-\frac{(\Lambda^2+m^2)^{3/2}}{6\pi}+\frac{|m|^3}{6\pi}\equiv A(m)+\frac{|m|^3}{6\pi}$, where a hard momentum cut-off $\Lambda$ is introduced. Physically, $\Lambda$ is order of surface band width. $A(m)$ is a cut-off dependent regular function which is $A(m)=-\frac{(\Lambda^2+m^2)^{3/2}}{6\pi}$ with the above simple regularization of a hard cut-off at $\Lambda$. In the following we show that general analytical expression can be obtained:
\begin{align}
 &\langle H_0+m\psi^{\dagger}\sigma_z\psi\rangle-A(m)\notag\\
=&\langle H_0+m\psi^{\dagger}\sigma_z\psi\rangle-\langle H_0+m\psi^{\dagger}\sigma_z\psi\rangle\lvert_{B=0}+\frac{|m|^{3}}{6\pi}\notag\\
=&-m\rho+\frac{B|m|}{2\pi}-\frac{(2B)^{3/2}}{4\pi}\zeta(-\frac{1}{2},\frac{m^2}{2B}),\label{eq:mf_energy}
\end{align}
which is another way to write Eq. (\ref{eq:vari_mf}). 

The coefficient of the second term of Eq. (\ref{eq:mf_energy}) is twice the previous incorrect result in Ref. \onlinecite{Zhukovsky:February2003:0040-5779:254}. The previous calculation basically sums over all filled LLs whose energies are given by Eq. (\ref{eq:ll_energy}), and then substract out an infinity to obtain the finite result. If this would be the correct regularization scheme one would expect that when $\nu=-1/2$ the energy is independent of the sign of $m$ because all the negative LLs' energy only depends on $m^2$. Interestingly, we will prove rather convincingly below that the energy function obtained in this way incorrect.

For any finite size lattice, the magnetic field only couples to hopping $t\rightarrow t e^{ia}$, indicating that the trace of the hamiltonian is invariant with or without magnetic field. 
This means that in the presence of particle-hole symmetry (which is true for both our lattice calculation below and the low energy effective theory in magnetic field), the summation of all energy levels remains zero independent of magnetic field. When magnetic field is turned on, the energy of the 0th LL is $-m$. This means that, the LLs with index $\pm n$ cannot have the same energy magnitude as shown in Eq. (\ref{eq:ll_energy}), otherwise the trace of the Hamiltonian would be $-m$. Let us  use $E_{-}(m)$ to denote the summation of the energies of all LLs with negative indices ($n=-1,-2,\cdots$) and use $E_{+}(m)$ to denote the summation of the energies of all LLs with positive indices ($n=1,2,\cdots$). The conservation of trace of Hamiltonian dictates: $E_{+}(m)+E_{-}(m)-m=0$ for arbitrary $B$. This means that at least one of $E_{+}(m)$ and $E_{-}(m)$ has linear term depending on $m$. In fact we find they each has a linear term $m/2$. Only when this linear shift is considered the regularization scheme can be correct.

As we mentioned, the correct regularization scheme is essential to obtain Eq. (\ref{eq:mf_energy}). One physically correct regularization is to treat the Dirac fermion as the continuum limit of a lattice model. We perform a numerical calculation of a lattice Dirac fermion model which confirms the validity of Eq. (\ref{eq:mf_energy}). Even thought it is true that it is impossible to realize a single gapless Dirac fermion in 2D by the no-go theorem, we can consider a quantum anomalous Hall mass in a lattice model with two gapless Dirac fermions, whose energy function is just twice of the single Dirac fermion case. Since the $\rho$ dependence in Eq. (\ref{eq:mf_energy}) is all from $-m\rho$, in the following we only need prove that Eq. (\ref{eq:mf_energy}) is correct for the case of $\rho=-\frac{B}{4\pi}$, i.e., $\nu=-1/2$ or a fully empty zeroth LL. 

We consider a time-reversal symmetric two-dimensional square lattice tight-binding model with two gapless Dirac cones\cite{hosur-2009}, and then put in the quantum anomalous Hall mass term as follows: 
\begin{align}
 &H_{\mathrm{latt.}}=\sum_k \Psi_k^{\dagger} 2t(\tau_x\sigma_y\sin k_x+\tau_x\sigma_x\sin k_y)\Psi_k\notag\\
&+\Psi_k^{\dagger}\big(a+b(\cos k_x+\cos k_y)\big)\tau_z\Psi_k + m\Psi_k^{\dagger}(-\sigma_z)\Psi_k.
\end{align}
In this model there are four spin-orbital coupled states per site: $\{ s_+,s_-,p_+,p_-\}$, and $\sigma$ and $\tau$ matrices are acting within the spin and orbital spaces respectively. Only on-site terms and nearest neighbor hoppings are considered. Let $m=0$ for the moment. We set $t=0.5$ and $a=-2b=0.5$ so that at half-filling there are two degenerate two-component Dirac cones at $\Gamma$ point with $v_F=1$. Now we turn on the quantum hall mass $m$. Depending on the sign of $m$ the 2D system is in either $\nu=1$ or $\nu=-1$ integer quantum hall states. We then turn on a orbital magnetic field $B$ by modifying $t\rightarrow te^{ia}$ and calculate the energy keeping the two 0th and higher LLs empty but lower LLs fully filled. By defining and computing 
\begin{eqnarray}
E_{\nu=-1/2}(m,B)\equiv\frac{1}{2}\big[\langle H_\mathrm{latt.} \rangle|_B-\langle H_\mathrm{latt.} \rangle|_{B=0}+\frac{|m|^3}{6\pi}\big]\nonumber\\
\end{eqnarray}
where the factor of $\frac{1}{2}$ is introduced because we should compute the contribution from only one node, we will show that it agrees with Eq. ~(\ref{eq:mf_energy}). 

\begin{figure} \includegraphics[width=0.4\textwidth]{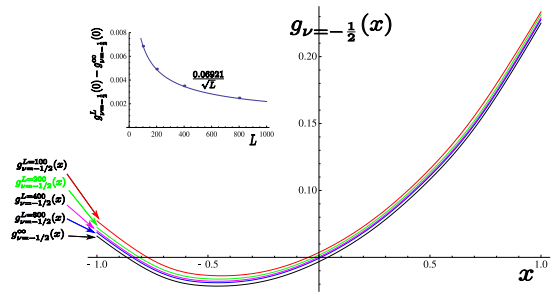}
\caption{Lattice calculation of various system sizes compared with $g^{\infty}_{\nu=-1/2}(x)$ based on Eq. (\ref{eq:mf_energy}). The inset studies the finite size scaling at $x=0$, which is consistent with the leading correction to scaling.}
\label{fig:lattice_calculation}
\end{figure}

We use periodic boundary condition and insert $L$ flux quanta uniformly over a sample of $L\times L$ lattice such that the magnetic field is $B=2\pi/L$. We expect that, at low energy, $E_{\nu=-1/2}(m,B)=g_{\nu=-1/2}(x)B^{3/2}$, where $x=m/\sqrt{B}$, by the scaling analysis. We numerically compute $g_{\nu=-1/2}(x)$ for various system sizes $L=100,~200,~400,~800$ as shown in Fig. \ref{fig:lattice_calculation}. If Eq. (\ref{eq:mf_energy}) is a correct form for energy function of a single Dirac node, we should be able to see that, as $L\to \infty$, $g_{\nu=-1/2}(x)$ should approach  $g^{\infty}_{\nu=-1/2}(x)\equiv \frac{x}{4\pi}+\frac{|x|}{2\pi}-\frac{2^{3/2}}{4\pi}\zeta (-\frac{1}{2}, \frac{x^2}{2}) $, the result based on Eq. (\ref{eq:mf_energy}). 

When $L$ goes to infinity, the numerical result from the lattice calculation indeed converges to the analytical result of Eq. (\ref{eq:mf_energy}), as shown in the Fig.\ref{fig:lattice_calculation}. The inset studies the finite size scaling where the deviation from thermodynamic result is nicely fit by $\propto 1/\sqrt{L}$. This finite size scaling is expected because when $x=m=0$, the energy scales as $\sim a B^{3/2}+b B^2 +...$ where the $B^2$ term is the leading correction to scaling. As the magnetic field that we used is $B=2\pi/L$, we expect the leading correction to scaling $\propto 1/\sqrt{L}$, which is consistent with the lattice numerical calculation.

Finally we mention that there are several simple limits of Eq. (\ref{eq:mf_energy}) that one can understand. When $m=0$ Eq.(\ref{eq:mf_energy}) becomes $\frac{\zeta(3/2)(2B)^{3/2}}{(4\pi)^2}$, which is a well-known result\cite{PhysRevD.29.2366,PhysRevLett.64.1166}. When $B=0$ the r.h.s. of Eq. (\ref{eq:mf_energy}) becomes $\frac{|m|^{3}}{6\pi}$ which is consistent with the second line. Note that the latter two terms depend on $m^2$ only: $[\frac{B|m|}{2\pi}-\frac{(2B)^{3/2}}{4\pi}\zeta(-\frac{1}{2},\frac{m^2}{2B})]\propto m^2$ for $m/\sqrt{B}\ll 1$ so that the first term of is the only term which couples $\rho$ and $m$ linearly.  The scaling behavior in Eq. (\ref{eq:mf_energy}): $B\sim\rho\sim m^{2}$ simply follows the fact that $1/m$ and $l_B$ are the only length scales when $U$ is absent.

\section{The large-$N$ analysis}
Now we include quantum fluctuations by a large-$N$ expansion. The mean-field study that we performed can be reinterpreted as the leading order of large-$N$ expansion of the following generalized model:
\begin{align}
 H=\psi_a^{\dagger}[\vec \Pi\cdot \vec\sigma]\psi_a
-G_{m}\frac{(\psi^{\dagger}_a\sigma_z\psi_a)^2}{2N} +G_{\rho}\frac{(\psi^{\dagger}_a\psi_a)^2}{2N},\label{eq:largeN}
\end{align}
where repeated indices means summation over $N$ flavors.
We tune $G_m$ to reach the quantum criticality $G_{m,c}$ in the large-$N$ limit, beyong which a mass $m\sum_{a=1}^{N}\psi^{\dagger}_a\sigma_z\psi_a$ term is generates dynamically and which is controlled by a Gross-Neveu universality. $G_{\rho}$ is irrelevant at this critical point and the free fermion fixed point and can be treated perturbatively. Quantum fluctuation is included if we consider a finite $N$. However whether the critical point $G_{m,c}$ is stable at the physical $N=1$ is unclear. It is worth to note that there are evidences of a stable $G_{m,c}$ at $N=1$ by functional renormalization group method\cite{PhysRevB.66.205111}. In our study we assume that $G_{m,c}$ is stable so that the transition at $U_c$ at zero magnetic field is continuous.

Now the relation between Hatree-Fock approximation and the large-$N$ analysis is clear. Without the $\frac{U\rho^2}{2}$ term, Eq.(\ref{eq:vari_mf}) would be identical to the effective action of Eq.(\ref{eq:largeN}) without the $G_\rho$ term, in the large-$N$ limit. The $\frac{U\rho^2}{2}$ term in Eq.(\ref{eq:vari_mf}), similar to the $G_\rho$ term in Eq.(\ref{eq:largeN}), is just a non-singular perturbation at the quantum critical point. 

We now show that, as shown in Fig. \ref{fig:phase_diagram}(a), phase separation occurs in regimes I and III  at least when $N$ is large. On the other hand, phase separation should occur in regime II even when $N=1$.

\begin{figure}
 \includegraphics[width=0.3\textwidth]{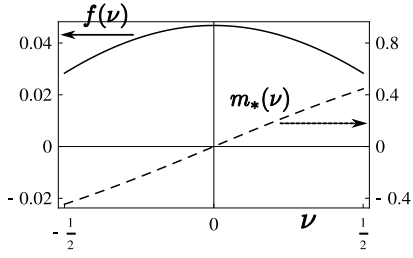}
\caption{We
plot the optimal mass $m_*(\nu)$ in units of $B^{1/2}$ and the
leading term of mean-field energy $H^{MF}_{sing}(\nu)=
f(\nu)B^{3/2}$ in the quantum critical regime. The non-convexity
signals phase separation.}
\label{fig:critical_phase_separation}
\end{figure}

First in the quantum critical regime I, the singular part of energy still have form $H_{sing}=f(\nu)B^{3/2}$ because energy has no anomalous dimension, which dominates any regular terms including $G_{\rho}B^2$ energy cost term. In Fig. \ref{fig:critical_phase_separation} we plot $f(\nu)$ in the large-$N$ limit (or at MF level). The non-convexity of $f(\nu)$ signals phase separation. At least when $N$ is large $f(\nu)$ is still non-convex and phase separation occurs. But when $N=1$ it is unclear whether this function is still non-convex. In fact the more interesting scenario is that when $N=1$, the ground state in regime I is in a homogenous gapped fractional quantum hall phase. If this is the case the fractional quantum hall phase can be a room temperature effect, because its excitation gap is comparable to the LL spacing.

Second in regime III, $l_B\gg \xi$ so the renormalization group flows to weak $U$ limit where $a m^2$ term is large and $m$ fluctuation is suppressed. Thus the phase separation energy gain $\propto B^2$ is reliable for any $N$. By scaling analysis $H_{gain}\sim \xi B^2$ where $\xi$ diverges as criticality is approached. It wins over any regular energy cost terms including $G_{\rho}B^2$. However when $N=1$ it is unclear whether $H_{gain}\sim \xi B^2$ is positive or negative.

Finally even when $N=1$ we have the following argument for phase separation to occur in regime II. In regime II the phase separation energy gain scales as $B$-linear (in particular when $U$ is close to $U_c$, energy density gain $\sim \frac{\hbar v_F}{l_B^2\xi}$ by hyper-scaling hypothesis.), which dominates any regular $B^2$ energy cost. Note that the $B$-linear dependence is reliable when quantum fluctuation is present, because at the length scale of $l_B$ the RG flows to the deeply ordered Gaussian FP where the mass magnitude $|m|$ is not fluctuating.

\section{Second order perturbation}
\begin{figure}
 \includegraphics[width=0.35\textwidth]{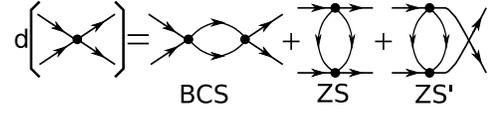}
\caption{The three processes of
the second order perturbation. Black dot represents $u$. External
legs are in the 0th LL and internal legs are in higher LLs.}
\label{fig:feynman_diagram}
\end{figure}

In the presence of magnetic field the single Dirac node form landau levels whose wave functions can be solved neatly.
\begin{align}
 &H_0=v_F \begin{pmatrix}
        0&\Pi_x-i\Pi_y\\
\Pi_x+i\Pi_y&0
       \end{pmatrix},&&\vec\Pi=\vec p+\frac{e}{c}\vec A.
\end{align}
If we define $a^{\dagger}=\frac{-il_B}{\sqrt{2}\hbar}(\Pi_x+i\Pi_y)$, then $[a,a^{\dagger}]=1$ and in this representation $H_0=\frac{\sqrt{2}v_F\hbar}{l_B}\begin{pmatrix}
        0&-ia\\
ia^{\dagger}&0
       \end{pmatrix}$. Apart from $a$, there is another set of variables describing the different states within the same landau level. Following MacDonald\cite{macdonald-1994}, one defines $b=\frac{1}{\sqrt{2}l_B}(z+i\frac{l_B^2}{\hbar}\Pi)$ where $z=x+iy$ and $\Pi=\Pi_x+i\Pi_y$. It is easy to show that $[b,b^{\dagger}]=1$ and $[a,b]=[a^{\dagger},b]=0$. These define a basis of wavefunctions: $|n,m\rangle=\frac{(a^{\dagger})^n(b^{\dagger})^m}{\sqrt{n!m!}}|0,0\rangle$ where $a|0,0\rangle=b|0,0\rangle=0$. And the energy levels and eigenfunctions are:
\begin{align}
|\Psi_{n,m}\rangle&=\frac{\sqrt{2}^{\delta_{n,0}}}{\sqrt{2}}\begin{bmatrix}-\mbox{sgn}(n)i|n-1,m\rangle\\|n,m\rangle\end{bmatrix},\notag\\
E_n&=\mbox{sgn}(n)\sqrt{2|n|}\cdot \frac{\hbar v_F}{l_B},
\end{align}
where $\mbox{sgn}(n)$ is the sign function and $\mbox{sgn}(0)=0$.

The full hamiltonian is $H=H_0+H_{int}$ where:
\begin{align}
 H_{int}=\sum_{i<j}\int \frac{dq^2}{(2\pi)^2}V(q)e^{iq\cdot (r_i-r_j)}.
\end{align}
For pure Coulomb interaction $V(q)=\frac{2\pi e^2}{\epsilon |q|}$. In the second quantization formulism:
\begin{align}
 H&=E_n\psi_{n,m}^{\dagger}\psi_{n,m}\notag\\
&+\frac{1}{2}\sum_{n_i,m_i}V_{4,3;2,1}\psi_{n_4,m_4}^{\dagger}\psi_{n_3,m_3}^{\dagger}\psi_{n_2,m_2}\psi_{n_1,m_1}
\end{align}
where $V_{4,3;2,1}$ is an abbreviation of:
\begin{align}
&V_{n_4,m_4,n_3,m_3;n_2,m_2,n_1,m_1}\notag\\
=&\hspace{-0.15cm}\int\hspace{-0.25cm}\frac{dq^2}{(2\pi)^2}\hspace{-0.05cm}V(q)\hspace{-0.05cm}\langle \Psi_{n_4,m_4}^i| \langle  \Psi_{n_3,m_3}^j|e^{iq\cdot(r_i-r_j)}|\Psi_{n_1,m_1} ^i\rangle|\Psi_{n_2,m_2}^j\rangle\notag\\
=&\hspace{-0.15cm}\int\hspace{-0.2cm}\frac{dq^2}{(2\pi)^2} V(q)e^{-|Q|^2}\frac{\sqrt{2}^{\sum_{i=1}^4\delta_{n_i,0}}}{4}G_{m_4,m_1}(-Q)G_{m_3,m_2}(Q)\notag\\
\cdot&\big[\mbox{sgn}(n_4)\mbox{sgn}(n_1)G_{|n_4|-1,|n_1|-1}(-\bar Q)+G_{|n_4|,|n_1|}(-\bar Q)\big]\notag\\
\cdot&\big[\mbox{sgn}(n_3)\mbox{sgn}(n_2)G_{|n_3|-1,|n_2|-1}(\bar Q)+G_{|n_3|,|n_2|}(\bar Q)\big].
\end{align}
Here $Q\equiv (q_x+iq_y)\cdot l_B$ and $G_{n_1,n_2}(Q)$ is related to generalized Laguerre polynomial $L_{n_2}^{n_1-n_2}$:
\begin{align}
 G_{n_1,n_2}(Q)=\left(\frac{n_2!}{n_1!}\right)^{\hspace{-0.1cm}\frac{1}{2}}\hspace{-0.1cm}\left(\frac{-iQ}{\sqrt{2}}\right)^{\hspace{-0.1cm}n_1-n_2}\hspace{-0.1cm}L_{n_2}^{n_1-n_2}\left(\frac{Q\bar Q}{2}\right).
\end{align}
Note that for Coulomb interaction $V_{4,3;2,1}\propto \frac{e^2}{\epsilon l_B}$.

Now we antisymmetrize $V_{4,3;2,1}$ by defining $\tilde V_{4,3;2,1}\equiv V_{4,3;2,1}-V_{3,4;2,1}$, and go to action formulism, $Z=\int \mathcal{D}\Psi^{\dagger} \mathcal{D}\Psi e^{S}$:
\begin{align}
 &S=\int \frac{d\omega}{2\pi}\sum_{n,m}\psi_{n,m}^{\dagger}(\omega)(i\omega-E_{n})\psi_{n,m}(\omega)\notag\\
&-\frac{1}{4}\int\prod_{i=1}^4\frac{d\omega_i}{2\pi}\sum_{n_i,m_i}u_{4,3;2,1}\psi_{n_4,m_4}^{\dagger}(\omega_4)\psi_{n_3,m_3}^{\dagger}(\omega_3)\notag\\
&\cdot\psi_{n_2,m_2}(\omega_2)\psi_{n_1,m_1}(\omega_1),
\end{align}
where $u_{4,3;2,1}\equiv \tilde V_{4,3;2,1}\delta^{\omega}_{4,3;2,1}$ and $\delta^{\omega}_{4,3;2,1}\equiv 2\pi\delta(\omega_4+\omega_3-\omega_2-\omega_1)$.

Next we integrate out fermion modes with $n\neq 0$ in a single step and obtain a renormalized interaction $u_{4,3;2,1}+du_{4,3;2,1}$ in the 0th LL. 
\begin{align}
 &d u_{4,3;2,1}\big|_{n_i=0}=\int\frac{d\omega_5 d\omega_6}{(2\pi)^2}\sum_{n_5,n_6\neq 0}^{m_5,m_6}\left[\frac{1}{2}u_{6,5;2,1}u_{4,3;6,5}\right.\notag\\
&\left.+u_{6,4;5,1}u_{3,5;2,6}
 -u_{6,3;5,1}u_{4,5;2,6}\right]\mathcal{G}_5\mathcal{G}_6,
\end{align}
where $\mathcal{G}_i=\frac{1}{i\omega_i-E_{n_i}}$. The three terms can be summarized in Fig.\ref{fig:feynman_diagram}. Using identity $\int\frac{d\omega}{2\pi}\frac{1}{(i\omega-A)(i\omega-B)}=\frac{\theta\big(\mbox{\footnotesize Re}(A)\big)-\theta\big(\mbox{\footnotesize Re}(B)\big)}{B-A}$ where $\theta(x)=1 \mbox{ if } x>0$ and $\theta(x)=0 \mbox{ otherwise}$, we perform the frequency integration:
\begin{align}
 &d u_{4,3;2,1}\big|_{n_i=0} \notag\\
=&\sum_{n_5,n_6\neq 0}^{m_5,m_6}\left[\hspace{-0.03cm}- \frac{1}{2}\tilde V_{6,5;2,1}\tilde V_{4,3;6,5}\frac{\theta(E_{n_6})-\theta(-E_{n_5})}{i(\omega_1+\omega_2)-E_{n_5}-E_{n_6}}\right.\notag\\
&+\tilde V_{6,4;5,1}\tilde V_{3,5;2,6}\frac{\theta(E_{n_5})-\theta(E_{n_6})}{i(\omega_2-\omega_3)+E_{n_6}-E_{n_5}}\notag\\
&\left.-\tilde V_{6,3;5,1}\tilde V_{4,5;2,6}\frac{\theta(E_{n_5})
-\theta(E_{n_6})}{i(\omega_2-\omega_4)+E_{n_6}-E_{n_5}}\right]\delta^{\omega}_{4,3;2,1}
\end{align}
The retardation effect gives the frequency dependence which can be neglected for low energy dynamics when relevant energy scales are much smaller than the energy gap, namely $\omega\ll E_1$. This allows us to go back to Hamiltonian formulism.

We finally present the renormalized hamiltonian in terms of the well-known Haldane's pseudopotentials. Because the interaction perserves the relative angular momentum of two particles in the same LL, one can parametrize the effective interaction as:
$V_m=\langle m, M_{CM}|H_{int}|m, M_{CM}\rangle$
where $m, M_{CM}$ are the relative and total angular momentum of the two particles respectively. The $\tilde V$  and peudopotentials $V_m$ are related by:
\begin{align}
 V_1\hspace{-0.05cm}&=\hspace{-0.05cm}\tilde V_{m_4=1,m_3=0;m_2=0,m_1=1}\notag\\
V_3\hspace{-0.05cm}&=\hspace{-0.05cm}\frac{1}{4}\big(\tilde V_{3,0;0,3}+3\tilde V_{2,1;1,2}-\sqrt{3}(\tilde V_{3,0;1,2}+\tilde V_{2,1;0,3})\big)
\end{align}
where in the second line we abbreviate the $m_i$ labels. We find for the 0th LL in units of $\frac{e^2}{\epsilon l_B}$:
\begin{align}
V_1&=V_1^{\mbox{\tiny Coulomb}}+dV_1=0.443-0.0068\alpha \notag\\
V_3&=V_3^{\mbox{\tiny Coulomb}}+dV_3=0.277+0.0028\alpha
\end{align}
If we trust this leading order perturbation result even for large $\alpha$, and using the $dV_1\sim dV_3$ phase diagram  obtained by exact diagonalization\cite{PhysRevLett.84.4685}, we find that when $0\leqslant\alpha\lesssim 4$ the system is in CFL phase, when $4\lesssim\alpha\lesssim 7$ the system is in Pfaffian phase and when $7\lesssim\alpha$ the system is in Stripe phase.

\section{The Zeeman effect}
\label{sec:zeeman}
The Zeeman coupling between electron spins and magnetic field is given by
\begin{eqnarray}
{\cal H}_{Z}=\psi^\dag \left[\frac{1}{2}g\mu_B B\sigma^z\right] \psi,
\end{eqnarray}
where $g$ is the $g$-factor. The Zeeman coupling simply renormalizes the mass term: $m\to \tilde m=m+\frac12 g\mu_B B$ in the energy function Eq. (\ref{eq:vari_mf}). By denoting $\tilde \rho=\rho+\frac12 g\mu_B B/U$,  the MF variational energy is given by
\begin{align}
&E(\tilde \rho,\tilde m;B)=-\tilde m\tilde \rho+\frac{B|\tilde m|}{2\pi} -\frac{(2B)^{3/2}}{4\pi}\zeta(-\frac{1}{2},\frac{\tilde m^2}{2B})\notag\\
& \qquad\qquad\qquad -\frac{(\tilde m^2+\Lambda^2)^{3/2}}{6\pi}
 +\frac{1}{2U}\tilde m^2+\frac{U\tilde \rho^2}{2},%\equiv E_{m} + E_{\rho},
\label{eq:re_energy}
\end{align}
which has the exactly same form as the equation without Zeeman coupling. Note that the difference is that $\langle\tilde \rho\rangle$ averaged spatially is $\tilde \rho_\ast= g\mu_B B/2U$ instead of $0$ for zero magnetic field. 

Suppose that Eq. (\ref{eq:re_energy}) is minimized by  an optimal mass is $\tilde m_\ast>0$ if the system is homogeneous namely $\tilde \rho=\tilde \rho_\ast$ everywhere.  To see if phase separation is favored, we consider the situation that the system macroscopically separates into two regions with the configuration of ($\tilde \rho=\tilde \rho_\ast +\delta \rho$, $\tilde m=\tilde m_\ast+\delta m$) and ($\tilde \rho=\tilde \rho_\ast -\delta \rho$, $\tilde m=\tilde  m_\ast-\delta m$) respectively with small $\delta \rho$ and $\delta m$. Then its average energy density is given by
\begin{eqnarray}
\langle E\rangle&=&[E(\tilde \rho_\ast+\delta \rho,\tilde m_\ast+\delta m) + E(\tilde \rho_\ast-\delta \rho,\tilde m_\ast-\delta m) ]/2\nonumber \\
&=&E(\tilde \rho_\ast,\tilde m_\ast;B)+\frac12 F''(\tilde m_\ast) \delta m^2 \nonumber\\
&&\qquad\qquad+ \frac{U}2(\delta \rho-\delta m/U)^2 +{\cal O}(\delta m^4),\label{E3}
\end{eqnarray}
where $F(\tilde m)$ is defined as
\begin{eqnarray}
F(\tilde m)=-\frac{(2B)^{3/2}}{4\pi}\zeta(-\frac{1}{2},\frac{\tilde m^2}{2B})-\frac{(\tilde m^2+\Lambda^2)^{3/2}}{6\pi}.
\end{eqnarray}
From Eq. (\ref{E3}), it is obvious that $\delta \rho=\delta m/U$ is required to minimize the energy function $\langle E\rangle$. Consequently, as long as $F''(\tilde m_\ast)<0$, phase separation will be favored. Explicitly, $F''(\tilde m_\ast)$ is given by
\begin{eqnarray}
F''(\tilde m_\ast)&=&-\frac{(2B)^{1/2}}{4\pi}\left[\zeta(\frac12,\frac{\tilde m_\ast^2}{2B})-\frac{\tilde m_\ast^2}{2B}\zeta(\frac32,\frac{\tilde m_\ast^2}{2B})\right]\nonumber\\
&&\qquad-\frac1{2\pi}[(\tilde m_\ast^2+\Lambda^2)^{1/2}+\frac{\tilde m_\ast^2}{(\tilde m_\ast^2+\Lambda^2)^{1/2}}], \nonumber\\
&=&\frac{(2B)^{1/2}}{4\pi}\Big[x\zeta(\frac32,x) -\zeta(\frac12,x)\nonumber\\
&&\qquad~~~-2(x+y)^{1/2}-2\frac{x}{(x+y)^{1/2}}\Big],\label{E5}
\end{eqnarray}
where $x=\tilde m_\ast^2/2B$ and $y=\Lambda^2/2B\sim l^2_B/a^2 \gg 1$ for currently available magnetic field. 
We know that $\tilde m_\ast\sim B$ since a non-zero $\tilde m_\ast$ is induced by magnetic field. Thus, $x=\tilde m_\ast^2/2B\sim  B$; for weak magnetic field, $x\ll 1$. Because $x\zeta(\frac32,x)-\zeta(\frac12,x)$ is a monotonically increasing function and its values $\sim [1.5, 4]$ for  $x\in [0,1]$, we expect $F''(\tilde m_\ast)\propto [{\cal O}(1)-2l_B/a]$ where ${\cal O}(1)$ is order of one.  
Consequently, the condition for $F''(\tilde m_\ast)$ to be negative namely tendency to phase separation is equivalent to $l_B\gg a$, which is always satisfied for currently available magnetic field. In other words, the Zeeman effect has no qualitative effect on the tendency of phase separation when only short-range interactions are considered. 

What happen when long-range Coulomb interaction is included?  What are the fates of the stripe phases in regime I and II predicted without Zeeman effect? Because Zeeman coupling does not give new length scale and can be viewed as an redefinition of $\nu$: $\nu\rightarrow\tilde\nu=\nu+\frac{\pi g\mu_B}{U}$. This means that the energy gain in the regime I still scales as $\sim B^{3/2}$ (it is a gain at least in the large-$N$ limit). So the predicted stripe phase in regime I is as robust as the case without Zeeman coupling. 

On the other hand, the ``axionic stripe'' phase in regime II may be destroyed by Zeeman effect. ``Axionic stripe'' phase requires that the phase separation energy density gain is linear in $B$, which in turn requires that the masses in the $\nu=1/2$ and $\nu=-1/2$ regions have opposite signs. As mentioned in the main text, the condition for this to happen is $\frac{4\pi\hbar g\mu_B}{e U}< 1$ ($U>U_c$ and $U_c\sim 2\pi\hbar v_F/ \Lambda$), otherwise both regions have the same mass sign so that the phase separation energy density gain scales as $B^2$ and phase separation will be destroyed by Coulomb interaction. 

Let us estimate the ratio $\frac{4\pi\hbar g\mu_B}{e U_c}$:
\begin{align}
\frac{4\pi\hbar g\mu_B}{e U_c}\sim \frac{4\pi\hbar g\frac{e\hbar}{2 m_e}\Lambda}{e 2\pi\hbar v_F}=g\frac{\hbar \Lambda}{m_e v_F}\sim g\frac{\lambda_e}{a},
\end{align}
where $\lambda_e$ is the de Broglie wavelength of the electron moving at $v_F$ and $a\sim 2\pi/\Lambda$ is the short distance cut-off which is comparable to the lattice spacing. This is only an order of magnitude estimate because, for instance, we use the mean-field value of $U_c$ and the cut-off $\Lambda$ cannot be accurately estimated. If we nevertheless plug in $v_F\sim 3\times 10^5$m/s (as in Bi$_2$Te$_3$) and $a=8$\AA, we find $\frac{4\pi\hbar g\mu_B}{e U_c}\sim 3g$, which is of order one. We thus expect in systems with large $v_F,a$ and small $g$, this ratio is smaller than one and the ``axionic stripe'' phase can be realized.

Finally we comment on the magnitude of $g$-factor for the surface states of TI. It is known that in TI systems with strong spin-orbital interaction, the $g$-factor in the bulk can be as large as 30$\sim$50 (for example, Bi$_2$Se$_3$ has $g\sim 30$ in the bulk\cite{BiSe_dielectric}.). If the $g$-factor on the surface is as large as in the bulk, the Zeeman coupling can be a large energy scale. However, it is not obvious that the $g$-factor on the surface is comparable with the one in the bulk. In fact, it is observed by STM experiments\cite{cheng-2010,Hanaguri-2010} that the 0th LL energy of the surface state is basically field independent up to $\sim 11$Tesla (the change of 0th LL energy is within a couple of meV as field is tuned from $3$Tesla to $11$Tesla). We conclude that the $g$-factor on the surface cannot be $\sim 30$, which would mean that the 0th LL energy shifts $\sim 15$meV if field is tuned from $3$Tesla to $11$Tesla. This is a direct evidence that the surface $g$-factor can be an order of magnitude smaller that in the bulk.

\bibliographystyle{apsrev}
\bibliography{/home/ranying/downloads/reference/simplifiedying}
\end{document}